\documentclass[%
 reprint,
 amsmath,amssymb,
pra,
]{revtex4-2}

\usepackage{graphicx}
\usepackage{dcolumn}
\usepackage{bm}

\usepackage{xcolor}
\usepackage{amsmath}
\usepackage{mathtools}
\usepackage{kbordermatrix}
\renewcommand{\kbldelim}{(}
\renewcommand{\kbrdelim}{)}
\usepackage{caption}
\usepackage{todonotes}
\setuptodonotes{inline} 
\usepackage{caption}
\usepackage{subcaption}

\usepackage[shortlabels]{enumitem}
\usepackage{dsfont}
\usepackage{amsmath}
\usepackage{amsthm}
\usepackage{mathtools}
\usepackage{amssymb}
\usepackage{amstext}
\usepackage{amsfonts}
\usepackage{nicefrac}

\renewcommand{\phi}{\varphi}

\newcommand{\R}{\mathbb{R}}

\renewcommand{\P}{\mathcal{P}}

\newcommand{\eq}{\hspace*{-0.15mm}=\hspace*{-0.15mm}}

\newtheorem{theorem}{Theorem}

\newtheorem*{theorem*}{Theorem}

\begin{document}

\preprint{APS/123-QED}

\title{Electronic wavefunction with maximally entangled MPS representation}

\author{Benedikt R. Graswald}
\email{benedikt.graswald@ma.tum.de }

\author{Gero Friesecke}%
 \email{gf@ma.tum.de}
\affiliation{%
  Department of Mathematics, Technical University of Munich, Germany
}%

\date{\today}

\begin{abstract}


We present an example of an electronic wavefunction with 
maximally entangled MPS representation, in the sense that the  
bond dimension is maximal and cannot be lowered by any re-ordering of the underlying one-body basis. Our construction works for any number of electrons and orbitals.
\end{abstract}

\maketitle


\section{Introduction}



It has long been recognized that matrix product states (MPS) yield accurate representations of quantum chemical wavefunctions. Such representations lie at the heart of the QC-DMRG method, a state-of-art method for strongly correlated systems \cite{white_martin, luoquin, legeza_roeder, review_legeza, scholl}. However, exactness requires exponentially large matrices with respect to the system size and the quality of the approximation is governed by the size of the discarded singular values of the corresponding unfoldings  $\psi^{\mu_1, \ldots,\mu_k}_{\mu_k+1,\ldots,\mu_L}$ \cite{scholl, hackbusch2012tensor}.

Unlike in spin chains with identical sites, where the required matrix sizes are connected to the entanglement between subsystems which is in turn governed by area laws \cite{plenio, eisert, verstrate, hastings}, the situation in quantum chemistry is more complicated. The role of the sites is then taken by the system's molecular orbitals, and the matrix ranks, the singular values, and the overall approximation quality is strongly influenced by the ordering of the orbitals \cite{BLMR, boguslawski, legeza_roeder, legeza-solyom, LiF}. As turns out, standard examples with maximal entanglement such as the fermionic Bell states (see below)  have the feature that the largest matrix rank (or bond dimension) for $L$ molecular orbitals occupied by $N= L/2$ electrons drops from maximal, $2^{L/2}$, to just 2 independently of $L$, under optimal re-ordering.  

Here we present an explicit, rather more intricately correlated state whose bond dimension stays at the maximal value $2^{L/2}$, regardless of any re-ordering.

\begin{figure}[h!]
    \centering
\begin{subfigure}[b]{0.4\textwidth}
         \centering
         \includegraphics[width=\textwidth]{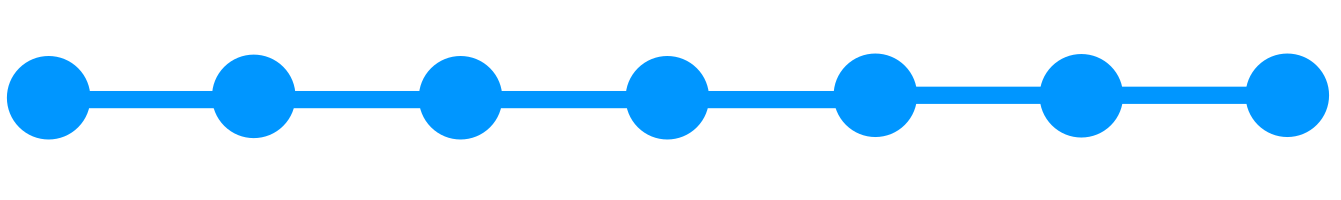}
         \caption{Original MPS}
\end{subfigure}
     \hfill
\begin{subfigure}[b]{0.4\textwidth}
         \centering
         \includegraphics[width=\textwidth]{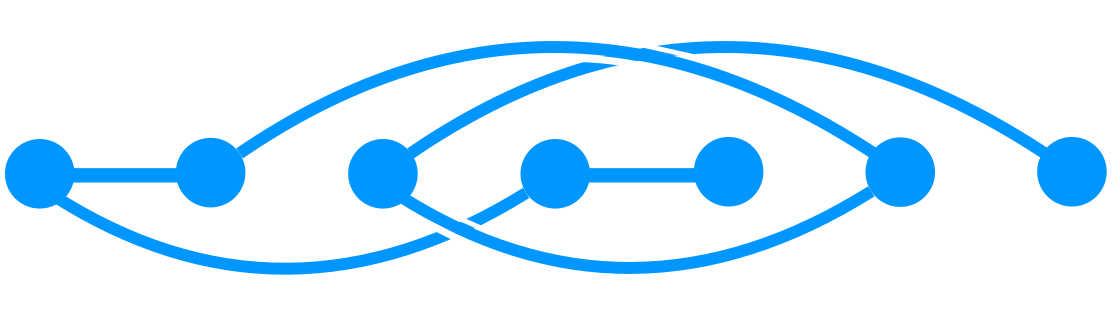}
         \caption{Reordered MPS}
\end{subfigure}
    \caption{A MPS depicted before and after reordering the orbitals (vertices). Bonds represent virtual variables, i.e.~summation indices in the matrix product \eqref{MPS}.}
    \label{fig:orderings}
\end{figure}

\section{MPS representation}

Given a suitable orthonormal set $\{\phi_1, \ldots,\phi_L\}$ of molecular spin orbitals, typically consisting of occupied and unoccupied Hartree-Fock orbitals, 
recall the exponential-sized full-CI expansion of an electronic wavefunction which reads, in $N$-particle respectively Fock space, 
\begin{equation}
\begin{aligned}
 \Psi &=
    \sum_{i_1 < \ldots<i_N} \lambda_{i_1,\ldots, i_N} 
    | \phi_{i_1}, \ldots, \phi_{i_N} \rangle \\
    &= \sum_{\mu_1, \ldots, \mu_L = 0}^{1} \psi_{\mu_1,\ldots, \mu_L } \Phi_{\mu_1,\ldots, \mu_L}.
\end{aligned}    
\end{equation}
Here the $|\varphi_{i_1},\ldots,\varphi_{i_N}\rangle$ are Slater determinants and
\begin{align} \nonumber
\psi_{\mu_1,\ldots, \mu_L} 
&= \begin{cases}
\lambda_{i_1,\ldots,i_N}, & \text{if }   \mu_{i_1}\eq \ldots \eq \mu_{i_N} \eq 1, \sum\limits_j \mu_j \eq N \\
0, & \text{ else,}
\end{cases} \\
\label{eq:occupation}
 \Phi_{\mu_1,\ldots, \mu_L} 
 &= \begin{cases}
| \phi_{i_1} \ldots, \phi_{i_N} \rangle,  & \text{if } \mu_{i_1} \eq \ldots \eq \mu_{i_N} =1 \\
0, & \text{ else.}
\end{cases}
\end{align}

The MPS approximation consists in the ansatz 
\begin{align} \label{MPS} 
  \psi_{\mu_1,\ldots, \mu_L} =  A_1[\mu_1]\, \cdots\, A_L[\mu_L],
\end{align}
where the $A_k[\mu_k]$ are matrices of size $1 \times M$ (for $k=1$), $M \times M$ (for $k=2, \ldots,L-1$), and $ M\times 1$ (for $k=L$) for some moderate value of $M$. 

\section{Fermionic Bell states}

Next, we argue that prototype examples of strong entanglement from spin physics and QIT -- like Bell states -- are in fact only weakly entangled in the MPS sense if re-ordering of the ``sites'' is allowed. Of course re-ordering only makes sense for molecular orbitals, not sites in 1D spin chains. 

For $N$ electrons occupying $L=2N$ orbitals $\{\varphi_1,\ldots,\varphi_L\}$, one can easily write down a fermionic analogon to the standard Bell states.

Set $\psi_k : =  \big( \phi_k + \phi_{k+N} \big)/\sqrt{2}$ for $k=1,\ldots,N$
and consider the Slater determinant $\Psi:=|\psi_1, \ldots, \psi_N \rangle$. It is then not hard to see (e.g.~\cite{dupuy2021inversion}) that its minimal MPS representation in the basis $\big(\phi_k\big)_k$ has bond dimension $2^N$.

Now apply a re-ordering which puts paired-up orbitals next to each other, 
\[
\big(\tilde{\phi}_1, \tilde{\phi}_2, \ldots ,\tilde{\phi}_{L-1}, \tilde{\phi}_{L} \big)
=
\big( \phi_1, \phi_{N+1}, \ldots, \phi_{N}, \phi_{L} \big).
\]
We claim that in the new basis $\big(\tilde{\phi}_k\big)_k$, $\Psi$ has an MPS representation with bond dimension just $2$. Indeed
\begin{align}
    \Psi &= \sum_{\mu_1, \ldots, \mu_L = 0}^{1} 
    A_1[\mu_1]\, \cdots\, A_L[\mu_L]\, \tilde{\Phi}_{\mu_1,\ldots, \mu_L}
\end{align}
where $\tilde{\Phi}$ is specified as in \eqref{eq:occupation} and the matrices $A_k$ are 
\begin{align*}
A_1[\mu_1] &= \begin{pmatrix} \delta^0_{\mu_1} & \delta^1_{\mu_1} \end{pmatrix}, ~A_L[\mu_L] =  \begin{pmatrix}\delta^1_{\mu_1}, \delta^0_{\mu_1} \end{pmatrix}^T, \\
A_{2\ell}[\mu_{2\ell}] &= \begin{pmatrix} \delta^1_{\mu_{2\ell}} & 0 \\ 0 & \delta^0_{\mu_{2\ell}}\end{pmatrix}, 
~A_{2\ell+1}[\mu_{2\ell+1}] = \begin{pmatrix} \delta^0_{\mu_{2\ell+1}} & \delta^1_{\mu_{2\ell+1}} \\ \delta^0_{\mu_{2\ell+1}} & \delta^1_{\mu_{2\ell+1}}\end{pmatrix}.
\end{align*}
Here $\ell = 1,\ldots, N-1$ and $\delta^{k}_\nu$ denotes the Kronecker delta.


\section{Maximally entangled state}

To construct a state whose bond dimension cannot be reduced by any re-ordering, we start off by recalling an old result by Besicovitch \cite{besicovitch1940linear}; let $p_1, \ldots p_s$, be different primes. Then
\begin{theorem}[Corollary 1 in \cite{besicovitch1940linear}]
A polynomial $P(\sqrt{p_1}, \ldots, \sqrt{p_s})$  with rational coefficients and degree w.r.t. each entry less than or equal to 1, not all equal to zero, cannot vanish.
\end{theorem}
Now we consider the set $\P:=\{ \sqrt{p_j}: p_j \text{ prime} \}$.
Then every matrix $A$ whose elements belong to $\P$ and are pairwise different has maximal rank, since -- for every square submatrix $B$ -- $\det(B)$ is exactly a polynomial of the above form.

Now define the state $\Psi_\P$ by
\begin{equation}
\begin{aligned} \label{eq:max_state}
    \Psi_\P
    &=
    \sum_{i_1 < \ldots<i_N} \lambda_{i_1,\ldots, i_N} 
    | \phi_{i_1}, \ldots, \phi_{i_N} \rangle \\ 
    &= \sum_{\mu_1, \ldots, \mu_L = 0}^{1} \psi_{\mu_1,\ldots, \mu_L } \Phi_{\mu_1,\ldots, \mu_L},
\end{aligned}    
\end{equation}
where the coefficients $\lambda_{i_1,\ldots,i_N}$ are mutually different elements of $\P$ and the second equation gives the occupation representation with $\psi_{\mu_1,\ldots, \mu_L}$ corresponding to  
$\lambda_{i_1,\ldots,i_N}$ as in \eqref{eq:occupation}. 
Then every unfolding $\psi^{\mu_{1}, \ldots,\mu_k}_{\mu_{k+1},\ldots, \mu_L}$ is a matrix of the above form and thus has maximal rank. 
In particular \cite{holtz2012manifolds}, $\Psi$ has maximal bond dimension.

Furthermore if we consider any new ordering, that is, we change our orbitals according to $(\phi_1,\ldots,\phi_L) = Q (\Tilde{\phi}_1, \ldots,\Tilde{\phi}_L)  $ with $Q \in \R^{L\times L}$ a permutation matrix, then we cannot decrease the rank of any unfolding.
Indeed, it is easy to see that we then obtain the following representation:
\begin{align*}
\Psi_\P &= \sum_{j_1<\ldots<j_N} \tilde{\lambda}_{j_1,\ldots,j_N} | \tilde{\phi}_{j_1}, \ldots, \tilde{\phi}_{j_N} \rangle \\
&=
\sum_{\mu_1, \ldots, \mu_L = 0}^{1} \tilde{\psi}_{\mu_1,\ldots, \mu_L } \tilde{\Phi}_{\mu_1,\ldots, \mu_L}
,
\end{align*}
with 
\[
\tilde{\lambda}_{j_1,\ldots,j_N} 
=
\sum_{i_1 < \ldots < i_N} \lambda_{i_1,\ldots,i_N} \begin{vmatrix}
q_{i_1,j_1} & \ldots & q_{i_1,j_N}\\
\vdots & & \vdots \\
q_{i_N,j_1} & \ldots & q_{i_N,j_N}
\end{vmatrix},
\]
where $q_{ij}$ denotes the elements of $Q$.
Since $Q$ is a permutation, exactly one determinant will be non-zero. Thus every unfolding still contains the same elements but only their positions change.
But by construction of the set $\P$, the position within the unfolding $ \tilde{\psi}^{\mu_{1}, \ldots,\mu_k}_{\mu_{k+1},\ldots, \mu_L}$ is irrelevant as long as all entries are different elements of $\P$. Hence the unfolding still has full rank. 
Therefore  $\Psi_\P$ still has maximal bond dimension.

We remark that in contrast to orderings, arbitrary fermionic mode transformations, i.e.~choosing the transformation $Q$ above as a unitary, can always somewhat decrease the bond dimension.
In the two-particle case ($N=2$) this can even achieve the optimal bond dimension of 3, for an arbitrary number of orbitals $L$ \cite{friesecke-graswald}.

\section{Singular value distribution}

We have also numerically calculated the singular value distribution of our example states for different values of $N$ and $L$ and different orderings (such as the widely used Fiedler order \cite{BLMR}) using the code tensor-train-julia \cite{dupuy_julia}.

\begin{figure}[ht!]
    \begin{center}
    \includegraphics[width = 0.5\textwidth, height=0.25\textheight]{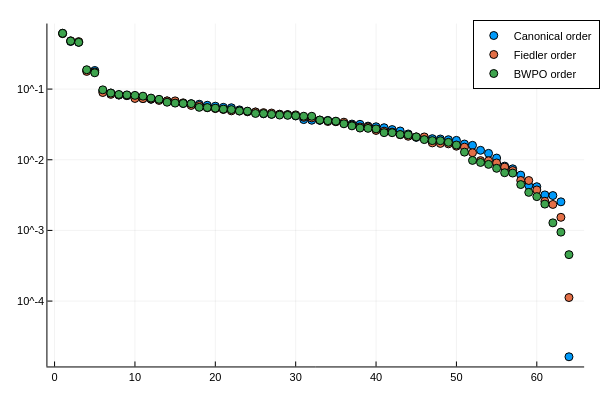}
    \end{center}
   \caption{Singular value distribution of the matrization $\psi^{\mu_1,\ldots,\mu_6}_{\mu_{7}, \ldots, \mu_{12}}$ of the state $\Psi_\P$ [eq.~\eqref{eq:max_state}] with $N=6$ electrons and $L=12$ orbitals,  
    for different orderings. 
    }
    \label{fig:singular_values}
\end{figure}
Fig.~\ref{fig:singular_values} corresponds to $N=6$, $L=12$, and a random choice of $\binom{L}{N}$ primes of size less than $2^{N+L}$. The different orderings shown are the original (canonical) order, the Fiedler order \cite{BLMR}, and the more recent best weighted prefactor order \cite{dupuy2021inversion}. In particular all $2^{L/2}$ singular values are non-zero, as predicted.

The singular values are seen to decay extremely slowly, and exhibit a remarkable almost-invariance under re-ordering. By contrast, for weakly correlated states re-ordering typically reduces the tail by several order of magnitude \cite{dupuy2021inversion}. 

Physically, the slow decay in Fig. \ref{fig:singular_values} means that for the state $\Psi_\P$, any two subsystems obtained by partitioning the molecular orbitals into two equal-size parts are strongly entangled.

\section*{Acknowledgements}
The authors thank M.-S. Dupuy for helpful discussions. 
Support from the International Research Training Group
IGDK Munich -- Graz funded by DFG, project number  188264188/GRK1754, is gratefully acknowledged.

\bibliography{apssamp}

\end{document}